\documentclass{aa}
\def\et{et al.}                          
\def\aap{A\&A}
\def\apj{ApJ}

\def\aj{AJ}
\def\aaps{A\&AS}

\def\baas{BAAS}
\def\sci{Sci}
\def\mnras{MNRAS}

\begin{document}

\thesaurus{05         
              (05.01.1;  
               03.20.2;  
               11.02.2 \object{BL\,1803+784}; 
               11.02.2 \object{BL\,2007+777}; 
               11.17.4 \object{QSO\,1928+738})}  

   \title{High precision difference astrometry applied to the triplet
of S5 radio sources \object{B1803+784}/\object{Q1928+738}/\object{B2007+777}}

   \author{E.\ Ros\thanks{{\it Present address:\/}
              Max-Planck Institut f\"ur Radioastronomie, Auf dem
              H\"ugel 69, D-53121 Bonn, Germany
              }\inst{1} 
\and J.M.\ Marcaide\inst{1}
\and J.C.\ Guirado\inst{1}
\and M.I.\ Ratner\inst{2}
\and I.I.\ Shapiro\inst{2}
\and T.P.\ Krichbaum\inst{3} 
\and A.\ Witzel\inst{3}
\and R.A.\ Preston\inst{4}
          }

   \offprints{E. Ros, ros@mpifr-bonn.mpg.de}

   \institute{Departament d'Astronomia i Astrof\'{\i}sica, 
              Universitat de Val\`encia,
              E-46100 Burjassot, Val\`encia, Spain
\and
              Harvard-Smithsonian Center for Astrophysics, 60 Garden
              St., Cambridge, MA 02138, US
\and
              Max-Planck Institut f\"ur Radioastronomie, Auf dem
              H\"ugel 69, D-53121 Bonn, Germany
\and
              Jet Propulsion Laboratory,
              California Institute of Technology, 4800 Oak Grove, 
              Pasadena, CA 91109, US
             }

   \date{Received 2 September 1998/ Accepted 6 April 1999}

   \maketitle

\markboth{E.\ Ros et al.: High precision 
astrometry on \object{B1803+784}/\object{Q1928+738}/\object{B2007+777}}{}

   \begin{abstract}

We determined
the separations
of the radio sources in the triangle
formed by the BL Lac objects {1803+784} 
and {2007+777}, and the \object{QSO\,1928+738}
from intercontinental interferometric observations carried 
out in November 1991 at the frequencies of 2.3 and 8.4\,GHz
simultaneously.
Difference phase-delay astrometry yielded
the following
separations:\\[1mm]
\begin{tabular}{r@{=}r@{}r@{}r@{}l@{$\pm$}l}
$\Delta \alpha_{\rm ({1803+784})-({1928+738})}$ &--1$^h$     & 27$^m$ & 2  & $\rlap{.}^s$811256 & 0$\rlap{.}^s$000062 \\
$\Delta \delta_{\rm ({1803+784})-({1928+738})}$ &  4$^\circ$ & 30$'$  & 2  & $\rlap{.}''$44833  & 0$\rlap{.}''$00012 \\ 
$\Delta \alpha_{\rm ({1928+738})-({2007+777})}$ &--0$^h$     & 37$^m$ & 42 & $\rlap{.}^s$503305 & 0$\rlap{.}^s$000033 \\
$\Delta \delta_{\rm ({1928+738})-({2007+777})}$ &--3$^\circ$ & 54$'$  & 41 & $\rlap{.}''$67756  & 0$\rlap{.}''$00013 \\
$\Delta \alpha_{\rm ({2007+777})-({1803+784})}$ &  2$^h$     & 4$^m$  & 45 & $\rlap{.}^s$314561 & 0$\rlap{.}^s$000052 \\
$\Delta \delta_{\rm ({2007+777})-({1803+784})}$ &--0$^\circ$ & 35$'$  & 20 & $\rlap{.}''$77077  & 0$\rlap{.}''$00013 \\
\end{tabular}
~\\[1mm]
We successfully connected differenced phase delays 
over 7$^\circ$ on the sky at 8.4\,GHz
at an epoch of maximum solar activity.  
The effects of the 
ionosphere on these VLBI data were mostly removed by estimates of
the total electron content from observations of GPS satellites.  
The comparison of the estimated separation of \object{QSO\,1928+738} and 
\object{BL\,2007+777}
with previous such estimates obtained from data at different epochs
leads us to a particular alignment of the
maps of \object{QSO\,1928+738} at different epochs
relative to those of \object{BL\,2007+777}, although with 
significant uncertainty.  For this alignment,
the jet components of \object{QSO\,1928+738} show a mean
proper motion of 0.32$\pm$0.10\,mas/yr 
and also suggest an identification for the
position of the core of this radio source.


\end{abstract}

%

%
%

\section{Introduction\label{introduction}}

A key need of astronomers is a celestial
reference frame.  Initially they used optical observations to obtain
such a reference frame (see, e.g., Fricke \et\ \cite{fri88},
Fricke \et\ \cite{fri91}).
A more rigid and accurate frame can be obtained using
Very Long Baseline Interferometry (VLBI), a technique that provides 
far higher angular resolution (see, for example, Johnston \et\ \cite{joh95}).
Using differenced phase delays
from observations of two sources
yields relative positions of radio sources with submilliarcsecond accuracy
and fractional precisions reaching $\sim$$2\cdot10^{-9}$
(Guirado \et\ \cite{gui95a,gui95b}, Lara \et\ \cite{lar96}).
Differenced phase delays may provide
the most precise observable in astrometry.
The group-delay observable, although also capable of providing
global astrometry,
has a statistical standard error larger than the 
phase-delay error by a factor equal to the ratio of the center frequency
of the VLBI observations to their effective bandwidth.

In radio astrometric work, it is important to account for
the propagation medium, including the ionosphere.
Sard\'on \et\ (\cite{sar94}) showed that the 
total electron content (TEC) of the ionosphere could be determined
with high accuracy (standard error of about 2\%) 
for all directions from a site by using
dual-frequency GPS data.
The ionospheric contribution can also be determined 
from dual-band VLBI
observations by comparing connected phase delays 
(or group delays) from the two bands.
In this paper, we compare those approaches for estimating
the effect of the ionosphere.

Our use of a triangle of targets allows
an improvement with respect to astrometry with pairs of radio sources
(e.g., Guirado \et\ \cite{gui95a,gui95b,gui98}):
closure on the sky provides a check on the internal consistency of
the data reduction.

%
%

\section{VLBI observations and data 
analysis\label{VLBI-observations-data-analysis}}

We observed the radio sources \object{BL\,1803+784}, \object{QSO\,1928+738}, and 
\object{BL\,2007+777} in right circular polarization at both 2.3 and
8.4\,GHz, on 20 and 21 November 1991
(MJD 48580 and 48581) from 14:00 to 04:00 UT.  Using a global VLBI array, we
recorded with the Mark III system (Rogers \et\ \cite{rog83}), in modes A (56\,MHz)
or B (28\,MHz), depending on the instrumentation at the individual sites.
The array consisted of the following
antennas (in parenthesis, code, diameter, location): 
Effelsberg (B, 100\,m, Germany); Hay\-stack (K, 37\,m,
Massachusetts, US); Medicina (L, 32\,m, Italy); DSS63 
(M, 70\,m, Spain); and phased-VLA (Y, 130\,m effective diameter, New
Mexico, US); Fort Davis (F, 25\,m, Texas, US); 
Kitt Peak (T, 25\,m, Arizona, US);  Pie
Town (P, 25\,m, New Mexico, US); Los Alamos (X, 25\,m, New Mexico, US); and 
Green Bank (G, 43\,m, West Virginia, US).  

The three radio sources were observed cyclically (cycle time of 9\,min,
clockwise direction) with
2 minutes ``on" radio source and 1 minute allowed for slewing antennas.
All sources had total flux densities greater than 1 Jy at the epoch 
of observation.  We
obtained strong detections from each of the baselines for at least 80\% of the
2 min ``scans".

The observations were plagued by problems.  
Medicina and Green Bank had phasing problems. DSS63 
recorded only 3 hours of data. The phased-VLA recorded only
8.4\,GHz
data, and Haystack was set to a wrong sky frequency.
After many fruitless attempts to use part of the data from these
stations, we decided to use only the data from the subset
of antennas with very good performance:
Effelsberg, Fort Davis, Pie Town, Kitt Peak, and Los Alamos.

The data were correlated at the Max-Planck Institut f\"ur
Radioastronomie, Bonn, Germany.  After fringe-fitting the correlator 
output, we obtained for each observation and
baseline, estimates of the group delay, 
the phase-delay rate, the fringe phase (modulo $2\pi$), and
the correlation amplitude at each of two selected reference frequencies 
(2,294.99 MHz and 8,404.99 MHz).
Following standard procedures we constructed the visibilities 
using the calibration information provided by staff of the antennas.
We made hybrid maps of each of the three radio sources at 2.3 and 8.4\,GHz
using the Caltech package (Pearson \cite{pea91}) and DIFMAP (Shepherd \et\
\cite{she95}).  
Our mapping results are presented in Sect.\ \ref{mapping-radio-sources}
and our astrometric results in Sects.\ \ref{astrometric-data-reduction} and
\ref{angular-separations}.

%
%

\section{Mapping of the radio sources \object{BL\,1803+784}, \object{QSO\,1928+738}, 
and \object{BL\,2007+777}\label{mapping-radio-sources}}

A complete sample (S5) (K\"uhr \et\ \cite{kuh81}) of compact radio 
sources has been studied using the VLBI technique 
over the last two decades (Eckart \et\ \cite{eck86,eck87}, 
Witzel \et\ \cite{wit88}). 
These sources were selected from the fifth installment of the MPIfR-NRAO
5\,GHz strong source survey.  The sources have flat spectra that
permit multiband observations.  All members of this sample
have neighbors relatively nearby on the sky, allowing phase-reference
observations to be made (Guirado \et\ \cite{gui95a,gui98}).  


\subsection{The BL Lac object 1803+784\label{bl-lac-object-1803}}

{1803+784} is a BL Lacertae type object with $z$=0.684 and $V$=16.4
(Stickel \et\ \cite{sti91}).  It exhibits a radio structure with 
kiloparsec-size
components (Antonucci \cite{ant86}, Strom \& Biermann \cite{str91}, 
Kollgaard \et\ \cite{kol92}, Murphy \et\ \cite{mur93}) to the south of the
core (at angular distances from the core of 2$''$, 
37$''$ and 45$''$).  On parsec scales 
the source shows a westward jet (Eckart \et\ \cite{eck86}, Pearson \& 
Readhead \cite{pea88}, Schalinski \cite{sch90});
VLBI monitoring of these milliarcsecond (mas)-size components 
at 22\,GHz shows components at angular distances of about 0.4, 1.4 and 5\,mas 
from the core (Eckart \et\ \cite{eck86}, Schalinski \cite{sch90}, Krichbaum
\et\ \cite{kri90}, \cite{kri93}, \cite{kri94a}, \cite{kri94b}).  
Observations at 22 and 43\,GHz also reveal other features that appear to be
moving between the core and the 1.4 mas
component.  
Steffen (\cite{ste94}) has modeled this radio source
as an adiabatically
expanding homogeneous plasma jet
moving along a helical trajectory.
The parsec-scale structure is misaligned with the
kiloparsec structure by $\sim$90$^\circ$.

\begin{figure}[htbp]
\vspace{110mm}
\includegraphics{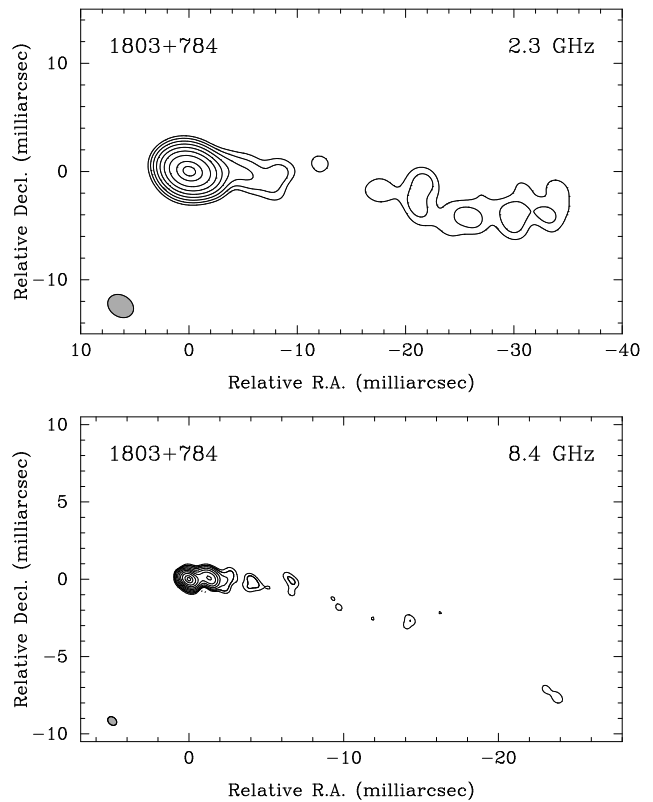}
\caption{Hybrid maps from observations on 20-21 November 1991 of 
the BL Lacertae object {1803+784} at 2.3 
and 8.4 GHz.  Contours are 0.5, 1, 2, 4, 8, 16, 32, 64,
and 90\% of the peak of brightness, 1.78 Jy/beam, for 2.3 GHz, and
{--0.25}, 0.25, 0.5, 1, 2, 4, 8, 16, 32, 64, and 90\% of 
the peak of brightness, 
1.40 Jy/beam, for 8.4 GHz. Dotted lines correspond to negative contours.
The size of the natural restoring beam, shown at 
the bottom left corner of each map, is 2.53$\times$1.97 mas (PA=58$^\circ$) 
for 2.3 GHz and 0.65$\times$0.49 mas (PA=49$^\circ$) for 8.4 GHz. 
Note that the scales of the two maps are not the same.
\label{fig-map-1803} }
\end{figure}

In our maps (see Fig.\ \ref{fig-map-1803}), we find well defined 
components to the west of the core, with
the jet structure extending up to 35\,mas from the core 
at 2.3\,GHz and up to 21\,mas at 8.4\,GHz.  The features in our maps agree
well with those shown in previous maps. 


\subsection{The \object{QSO\,1928+738} (\object{4C\,73.18})\label{qso-1928}}

\object{QSO\,1928+738}, also known as \object{4C\,73.18}, is a well-studied
quasar with $z$=0.302 and $V$=15.5 (Lawrence \et\ \cite{law86}).  
On kiloparsec scales, it exhibits two-sided jet
emission feeding two symmetrically placed lobes in a north-south
orientation (Rusk \& Rusk \cite{rus86}, 
Johnston \et\ \cite{joh87}, Murphy \et\ \cite{mur93}).  
On parsec scales it
displays a pronounced one-sided southward jet with superluminal motion
along a sinusoidally curved jet 
(Eckart \cite{eck85}, Pearson \& Readhead \cite{pea88}, Hummel \cite{hum90}, 
and Hummel \et\ 1992, hereafter \cite{hum92}).  
\cite{hum92} describe this source as having a helical magnetic field 
attached to a rotating accretion disk.
Further, Roos \et\ (\cite{roo93}) 
suggest that the source consists of two massive black holes, 
with mass ratio smaller than 10, total mass $\sim$10$^{8}M_\odot$
and orbital radius $\sim$10$^{16}$cm.
Such massive binary black holes could be
responsible for the sinusoidal jet ridge line which has a period of
precession of about 2.9 years according to the ballistic relativistic
jet model.  

\begin{figure*}[bhtp]
\vspace{84mm}
\includegraphics{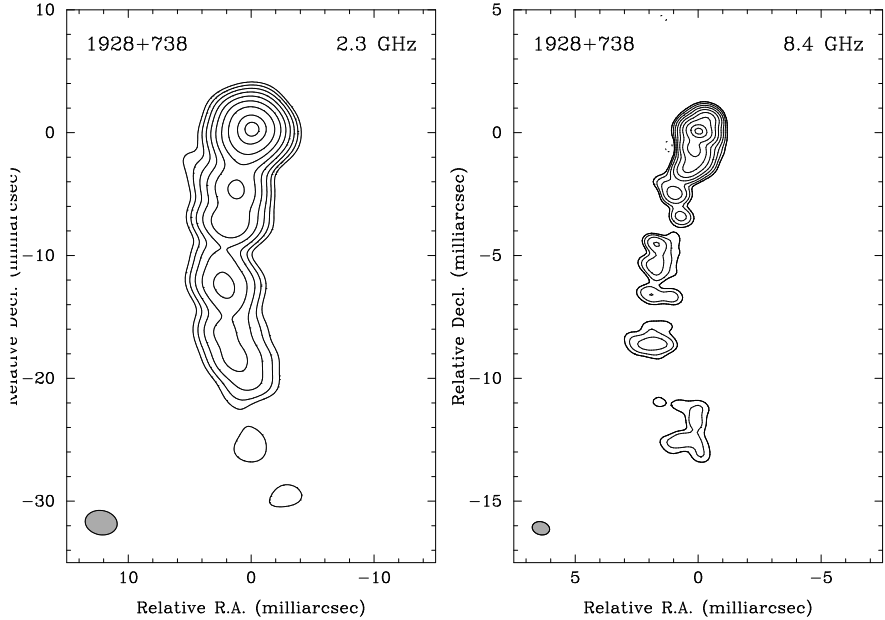}
\vspace{-70mm}
\hfill      
\parbox[b]{5cm}{\caption[]{
Hybrid maps from observations on 20-21 November 1991
of the \object{QSO\,1928+738} (\object{4C\,73.18}) at 2.3 and 8.4 GHz.
Contours are 0.5, 1, 2, 4, 8, 16, 32, 64, and 90\% of 
the peak of brightness, 1.16 Jy/beam, for 2.3 GHz, and
--0.5, 0.5, 1, 2, 4, 8, 16, 32, 64, and 90\% of the peak of brightness, 1.18
Jy/beam, for 8.4 GHz.  Dotted lines correspond to negative contours.
The size of the natural restoring beam, shown at the bottom left
corner of each map, is 1.98$\times$2.62 mas (PA=80$^\circ$) for 2.3 GHz and 
0.72$\times$0.53 mas (PA=76$^\circ$) for 8.4 GHz.
Note that the scales of the two maps are not the same. 
\label{fig-map-1928} }
}
\end{figure*}

Our maps (see Fig.\ \ref{fig-map-1928}) show the core-jet 
structure, with components in the jet
up to 30 mas south of the core at 2.3\,GHz and up to 
13\,mas south at 8.4\,GHz.  

\subsection{The BL Lac object {2007+777}\label{bl-lac-object-2007}}

2007+777 is also a BL Lacertae type object with $z$=0.342 and $V$=16.5
(Stickel \et\ \cite{sti91}).  It is a core-dominated radio source with a 
westward-directed jet of parsec scale with superluminal
components discernible up to 6\,mas from the core (Eckart \cite{eck83}).
This source has also been studied on a kiloparsec scale (Antonucci \cite{ant86},
Kollgaard \et\ \cite{kol92}).  
The kiloparsec structure is aligned with the parsec structure and has
components $\sim$11$''$ east of the core, and a jetlike feature
extending 16$''$ to the west with components at $\sim$8.5$''$ and
$\sim$15.8$''$ from the core.  

\begin{figure}[htbp]
\vspace{115mm}
\includegraphics{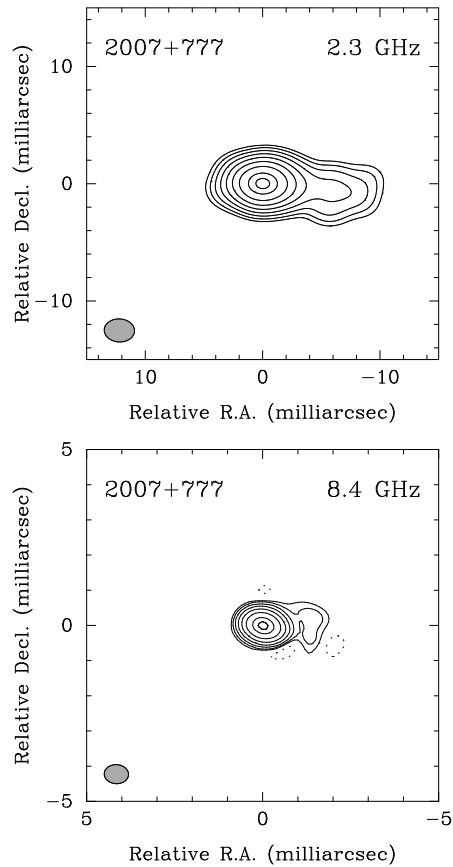}
\caption{Hybrid maps from observations on 20-21 November 1991
of the BL Lacertae object {2007+777} at 2.3 and 8.4 GHz.
Contours are 0.5, 1, 2, 4, 8, 16, 32, 64, and 90\% of 
the peak of brightness, 0.69 Jy/beam, for 2.3 GHz, and
--1, 1, 2, 4, 8, 16, 32, 64, and 90\% of the peak of brightness, 2.06
Jy/beam, for 8.4 GHz.  Dotted lines correspond to negative contours.  
The size of the natural restoring beam, shown at the bottom left
corner of each map, is  2.58$\times$1.97 mas (PA=87$^\circ$) for 2.3 GHz and 
0.69$\times$0.54 mas (PA=87$^\circ$) for 8.4 GHz.
Note that the scales of the two maps are not the same. 
\label{fig-map-2007} }
\end{figure}

For this source our maps (see Fig.\ \ref{fig-map-2007}) show a 
relatively small jet extending 8\,mas west of the core. We identify the 
brightest component of the map
as the core due to 
its spectral power being the same at 
2.3\,GHz and 8.4\,GHz.

%
%

\section{Astrometric data reduction\label{astrometric-data-reduction}}

Our astrometric data reduction contains several steps which yield
nearly ``ionosphere-free" and ``struc\-ture-free" phase-connected delays.
Using an extensively modernized version of software originally written
by Robertson (\cite{rob75}), we analyze these delays via weighted 
least-squares to estimate the angular separations among the sources.
In the next paragraphs we discuss in detail each of the steps of this
process.


\subsection{Phase connection\label{phase-connection}}

We used the
8.4\,GHz group delays, $\tau_g$, together with the 
phase-delay rates, ${\dot{\tau}}_\phi$, to 
estimate via weighted-least-squares the values of the 
parameters of our theoretical models
including the behavior of the atomic clock at each site relative
to that at a reference site.  
Using this astrometric 
model
we predicted for each of our ten 
baselines the number of cycles of phase between successive observations 
of the same source to permit us to ``connect" the phase delays
(Shapiro \et\ \cite{sha79}).
Denoting the reference frequency (8,404.99 MHz) by $\nu_r$ 
and the cycle time of the observations (9\,min)
by $\Delta t$, we can represent the constraint on the standard error,
$\sigma(\dot{\tau}_{\phi})$, needed to insure proper phase
connection: $\sigma(\dot{\tau}_{\phi}) \ll (\nu_r \Delta t)^{-1}$, i.e., 
$\sigma(\dot{\tau}_{\phi})$ must be small compared to
0.2\,ps$\cdot$s$^{-1}$. 
In our observations, this criterion is satisfied
at 8.4\,GHz for our observations and essentially complete phase 
connection was feasible.
However, phase connection was not possible for the data at 2.3\,GHz, due to
the strong effect of the ionosphere and the consequent
high scatter in the phase-delay data, leading to a violation of the 
criterion stated above.

%
%
\begin{table*}[htbp]
\caption{Parameters of the theoretical
model used in the weighted-least-squares analysis. \label{table-model} }
\begin{flushleft}
\begin{tabular}{lll} \hline
\multicolumn{3}{l}{{\bf Source coordinates (J2000.0)$^{\rm a}$}}  \\
{1928+738}$^{\rm b}$ &  $\alpha = 19^{h} 27^{m}  48\rlap{.}^s 495213$ & $\delta = 73^{\circ} 58' 01\rlap{.}''56995$ \\
\end{tabular}

\begin{tabular}{lrrrlcc}
\multicolumn{7}{l}{\bf Antenna site coordinates and propagation medium parameters$^{\rm a,c}$} \\
{\small Radio}
  &              &              &             & \multicolumn{1}{c}{\small Axis} 
						       & \multicolumn{2}{c}{\small Mean tropospheric} \\
{\small telescopes$^{\rm d}$} 
  & \multicolumn{3}{c} {\small Cartesian coordinates [m]} 
					      & \multicolumn{1}{c}{\small offset$^{\rm e}$ \,[m]} 
						       & \multicolumn{2}{c}{\small zenith delay$^{\rm f}$ [ns]}\\
  & \multicolumn{1}{c}{X} 
		 & \multicolumn{1}{c}{Y} 
				& \multicolumn{1}{c}{Z} 
					      &        &  $\tau_{\rm dry}$ 
 							     & $\tau_{\rm wet}$\\
B & 4033947.528  &   486990.428 & 4900430.727 & 0.0    & 7.4 & 0.2 \\
F &--1324009.072 &--5332181.961 & 3231962.515 & 2.123  & 6.4 & 0.4 \\
P &--1640953.650 &--5014816.009 & 3575411.928 & 2.136  & 5.9 & 0.1 \\
T &--1995678.575 &--5037317.707 & 3357328.183 & 2.138  & 6.2 & 0.0 \\
X &--1449752.303 &--4975298.589 & 3709123.975 & 2.139  & 6.2 & 0.0 
\end{tabular}

\begin{tabular}{lll}
{\bf Earth Tides}  &  & \\
Radial Love number, {\bf h}=0.60967 
                                  &  Horizontal Love number, {\bf l}=0.085 
                                                                       & Tidal lag angle, $\theta$ = 0$\rlap{.}^\circ$0 \\[1mm]
{\bf Earth Orientation Parameters$^{\rm a}$} &  & \\
{\bf Nutation}                    & {\bf Polar Motion and UT1}         & {\bf Precession Constant (J2000.0)$^{\rm g}$} \\
IAU 1980 Nutation Series +        & Wobble: {\bf x}=$0\rlap{.}''25221$ {\bf y}=$0\rlap{.}''27880$ 
                                                                                  & p=$5029\rlap{.}''.0966$/Julian century \\
IERS corrections for MJD=$48580$: & IERS LOD correction$=0.002422s$    & {\bf Mean obliquity (J2000.0)$^{\rm g}$}\\    
$d\psi=-0\rlap{.}''01238 ~~~d\epsilon=-0\rlap{.}''00201$     
                                             & UT1-UTC$=-0.027891s$    & $\epsilon_0$=23$^\circ26'21\rlap{.}''$448=84381$\rlap{.}''$448 \\
\hline
\end{tabular} 
\begin{list}{}{
\setlength{\leftmargin}{0pt}
\setlength{\rightmargin}{0pt}
}
\item[$^{\rm a}$] The reference coordinates
are taken from 1993 IERS Annual Report (IERS \cite{ier94}); they
are used only for convenience.  
We used standard deviations from
the IERS parameter estimates, but increased them twofold 
to be conservative.  These increased standard deviations are:  
0.035\,ms and 0.28\,mas for the $(\alpha,\delta)$ \object{QSO\,1928+738} coordinates,
2\,cm in each coordinate of site position,
1\,mas in each pole coordinate, and 0.04\,ms in UT1-UTC.  
See Table~\ref{table-sens}.
\item[$^{\rm b}$] The position of \object{QSO\,1928+738} was held fixed and the positions
of the other two sources were determined with respect to it.
\item[$^{\rm c}$] The site positions were
calculated by linear extrapolation (using the constant velocities for the
coordinates obtained from IERS (\cite{ier94}) 
and applied to its 1993.0 positions).
\item[$^{\rm d}$] See text for key.
\item[$^{\rm e}$] All the antennas are altitude-azimuth.
\item[$^{\rm f}$] Mean values of the zenith delays, computed using the
meteorological values provided by the staff of the observing antennas
(see, e.g., Saastamoinen \cite{saa73}).
\item[$^{\rm g}$] Lieske \et\ (\cite{lie77}).
\end{list}
\end{flushleft}
\end{table*}

We also considered possible overall offsets (equivalent to an integral
number of cycles of phase for all the baselines involving a given site and 
source, i.e., constant clock offsets or fringe ambiguity offsets).  To estimate 
such possible offsets, we considered the fringe ambiguity offsets as 
parameters whose values were to be estimated and we
introduced one of them into 
the global weighted-least-squares analysis
for each site for each radio source.  The values obtained were all
less than 24\,ps, a fifth of an ambiguity interval.  
We concluded that the fringe ambiguity offsets 
for the phase-delay data at 8.4\,GHz are all zero. 


\subsection{Radio source structure 
correction\label{radio-source-structure-correction}}

The selection of the reference point in each map is a critical
task.  The meaningful accuracy of our final sky-position estimate
can be no 
better than the accuracy with which we can identify a specific ``physical"
reference point in our 
map.  Using the Caltech package 
(Pearson \cite{pea91}), we made special renditions 
of the images of our radio sources with a pixel size of 0.01 mas for 8.4\,GHz 
(about 70 times smaller than the size of the CLEAN beam).  
For Figs.\ \ref{fig-map-1803}--\ref{fig-map-2007}, the
delta-function components from the CLEAN procedure had been
convolved with the corresponding Gaussian beams.  For these
special renditions of the images we also convolved
the CLEAN components with their corresponding Gaussian beams.  We chose
as the reference point the center of the pixel corresponding to the peak
of the brightness distribution, for each radio source and frequency.
For a source with 
a ``sharp" peak, the standard error in this location is proportional to
the ratio between the size of the interferometric beam  
and the SNR of the peak of brightness in the map (see, for example,
Fomalont \cite{fom89}). 
Using our reference points, we
calculated the structure contributions to the phase delay.
None of the contributions is larger than 25\,ps.
Thus, the structure contributions at 8.4\,GHz are 
$\sim$0.2\,cycles (no more than 70$^\circ$)
for \object{QSO\,1928+738}, the source with the largest structure.

Another choice (see Guirado \et\ \cite{gui95a}) of reference point
is the centroid of those
delta-function components with flux densities larger than 25\% of the
component with the highest flux density, in the area
covered by the ``main lobe" of the
synthesized beam.  The difference in the positions
determined from these two criteria is $\le$\,20\,$\mu$as.


\subsection{Troposphere correction\label{troposphere-correction}}

The 
neutral atmosphere introduces
a delay $\tau_{\rm atm}$ in the optical path 
length of the radio wave of about 2.3\,m in the
zenith direction.  From meteorological data obtained at each antenna site, we
calculated (see, e.g., Saastamoinen \cite{saa73}) a priori values for the two 
components of the delay at local zenith: the dry and the wet.  We show 
these values in Table~\ref{table-model}.

For the weighted-least-squares analysis, we represented
the zenith delay as a continuous, piecewise-linear 
function of time and determined it from meteorological values sampled 
every two hours during the observation.
We used the Chao mapping function (Chao \cite{cha72,cha74}) 
to approximate the delay at elevation angles other than the zenith.
These angles were
always greater than 20$^{\circ}$, and for these delays
the Chao mapping function yields results consistent with those from
other more accurate mapping functions.


\subsection{Ionosphere correction\label{ionosphere-correction}}

The plasma delay due to the Earth's ionosphere was 
the most difficult contribution to estimate.
With observations made at two frequency
bands simultaneously, the ionospheric contribution 
can largely be removed from some observations by taking
advantage of the $\nu^{-2}$ dependence of the effect.  
The delay of a radio wave propagating through a (weak) plasma
can be approximated by: $\Delta\tau_{ion}=\pm(k E_c)/(c \nu^2)$ (see,
for example, Thompson \et\ \cite{tho86}), where $k$=40.3
m$^3$s$^{-2}$, $c$ is the speed of light (m$\cdot$s$^{-1}$),
$\nu$ the frequency (Hz), and $E_c$ the integrated electron content 
(m$^{-2}$) along the line of sight; the plus and minus signs refer to
group and phase delays, respectively.
Unfortunately, for our observations, which were made near
a maximum in solar activity, we could not phase connect the delays 
at 2.3\,GHz, as noted in Sect.\ \ref{phase-connection}.  
Although the group-delay data at 2.3\,GHz and at 8.4\,GHz can be 
combined to yield ionospheric corrections, these are not accurate enough
to allow us to phase connect the 2.3\,GHz delay data.

Because we believed the corrections for the ionosphere would be somewhat
more accurate, we used estimates of 
$E_c$ from Global Positioning System (GPS) data
(Sard\'on \et\ \cite{sar94}) instead of from the dual-band VLBI data.  
The GPS system uses 
two frequency bands, one each at 1,575.42\,MHz
and 1,227.60\,MHz.  From data obtained simultaneously at one site
from different satellites, we estimated the total electron
content in the zenith direction.  With such estimates from a number of
sites, together with our model of the electron distribution in the ionosphere, 
we can interpolate to obtain the delay in a given direction for a given
(relatively nearby) site (see Klobuchar \cite{klo75}).
We therefore obtained data from several GPS stations 
for the time span of our observations.  
For the U.S.\ part of the array
we used the TEC data from Goldstone (CA,
US, 116.890$^\circ$W, 35.426$^\circ$N) and Pinyon Flats (CA, US,
116.458$^\circ$W, 33.612$^\circ$N) available at the GARNER archives
at SOPAC, University of California-San Diego,
to estimate the TEC for each of our VLBI observations
for the following antennas: 
Fort Davis (103.944$^\circ$W, 30.635$^\circ$N), 
Pie Town (108.119$^\circ$W, 34.301$^\circ$N),
Kitt Peak (111.612$^\circ$W, 31.956$^\circ$N),
and Los Alamos (106.245$^\circ$W, 35.775$^\circ$N), in each case
introducing a longitude (time) correction and an appropriate ``slant"
factor.  In our relatively crude procedure, following the geometry from
Klobuchar (1975), we assumed that the TEC was
concentrated at an altitude of 350\,km.
To determine the TEC for Effelsberg
(6.844$^\circ$E, 50.336$^\circ$N),
we used the
data from Herstmonceux (UK, 0.336$^\circ$E, 50.867$^\circ$N) and 
Wettzell (Koetzling, Germany, 12.789$^\circ$E, 49.144$^\circ$N) obtained
from the archives mentioned above.
Further details will be provided 
in Ros \et\ (\cite{ros99}).
We used the GPS results as
described above to deduce ionosphere corrections for the line of sight
from each VLBI station to each radio source for each 
scan, and to thereby correct
the group delays (at both frequencies) and phase delays (at 
8.4\,GHz).
The corrections were as large as 10 turns (cycles) at
8.4\,GHz for some scans on intercontinental
baselines.  The statistical standard errors of the GPS corrections
at 8.4\,GHz are 30\,ps, a fourth of
an ambiguity interval, corresponding to a TEC standard error of
1.5$\times$10$^{16}$\,el/m$^2$.


\subsection{Differenced observables
\label{differenced-observables}}

After the phase-connected observables at 8.4\,GHz
for each source were corrected for
source structure and ionospheric
effects, we formed the differenced
phase delays by subtracting the phase delay of each observation of a
radio source of the triangle from the previous one observed in the
cycle. 
The differenced phase delays are thus partially free from
unmodeled effects that remain in the phase delays for each one of the
sources, such as those effects due to instabilities of the time standards used
at the sites and inadequacies in the modeling of the troposphere.
When working with sources nearby to one another on the sky, subtraction
of results from consecutive observations tends to cancel errors since
the received signals from the two sources follow similar paths at nearly
the same epochs.  With
sources at larger angular separations, such tendencies to cancel are
weaker: not only are the spatial separations of the ray paths larger, but
the epochs of successive observations tend to be further apart, due to
the limited slew speeds of the antennas.  In our case the source 
separations range from 4$^\circ$ to 7$^\circ$.

Next, the undifferenced and differenced
phase-delay data from the three radio sources were
used in a weighted-least-squares analysis
(the use of different subsets of data, e.g., undifferenced
data from only one or two of the three radio sources, 
or the use of two or three sets of differenced data
does not alter the results or the statistical standard errors 
significantly).
We included undifferenced observables to better estimate the
long-term behavior of 
the station clocks.
For each baseline, we scaled separately
the standard deviations 
of the undifferenced phase delays for each radio source  and
the differenced phase delays for each pair
of radio sources 
to give a root-weighted-mean-square of unity
for the corresponding set of postfit residuals.
The resulting undifferenced phase delays have about threefold larger standard
errors than do the differenced phase delays.  

At this stage of the data reduction we again checked for
possible nonzero fringe ambiguity 
offsets and possible errors in phase connection. 
Estimating the fringe ambiguity
offsets in a least-squares analysis again yielded small
values (a few picoseconds), showing that no changes were necessary.

%
%

\section{Angular separations\label{angular-separations}}

\begin{figure*}[htbp]
\vspace{151mm}
\includegraphics{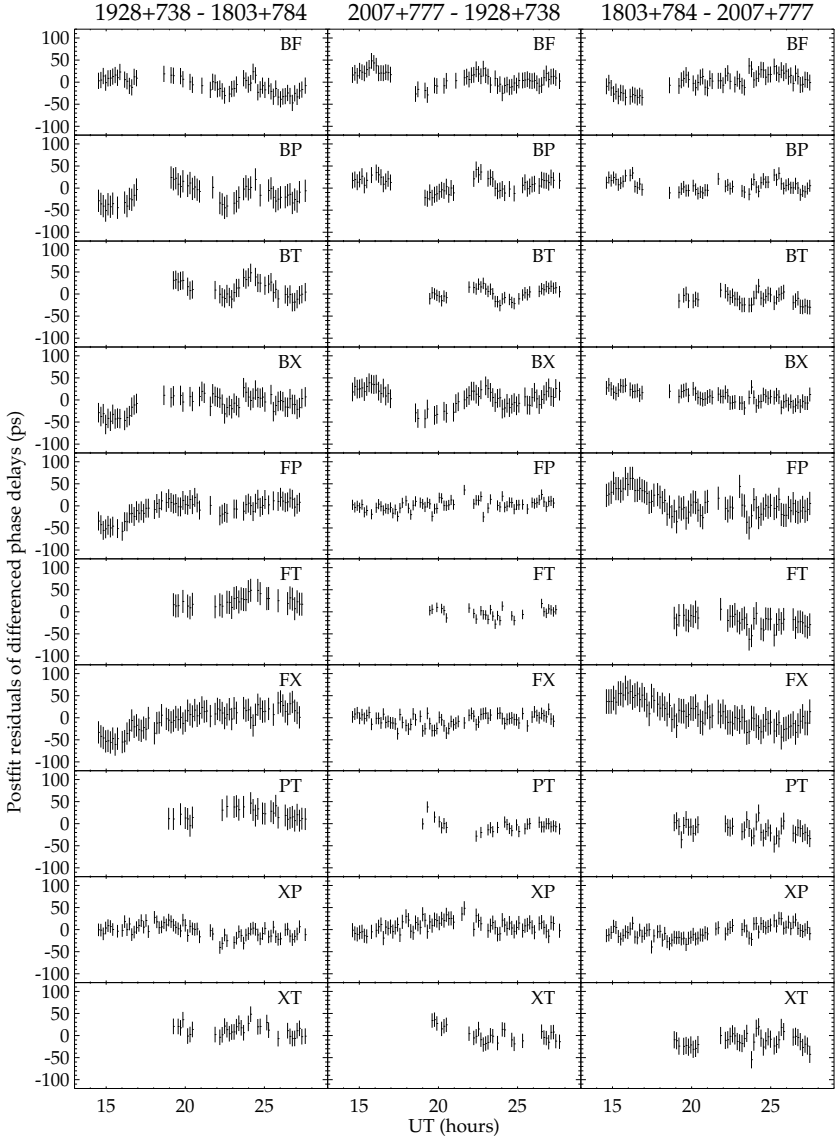}
\vspace{-3cm}
\hfill      \parbox[b]{5cm}{\caption[]{
Postfit residuals of the differenced phase delays for
all baselines and source pairs involved in our analysis (see text).
Full (vertical) scale is $\pm$ one phase cycle ($\pm2\pi$ 
equivalent), i.e., $\pm$119\,ps.  \label{fig-diff-x} }
}
\end{figure*}

With these phase-connected, nearly ionosphere- and struc\-ture-free
phase-delay data, we estimated the angular separations among the pairs
of radio sources of the triangle.  Specifically, we
estimated the coordinates of \object{BL\,1803+784} and \object{BL\,2007+777}
with respect to the fixed coordinates of \object{QSO\,1928+738}.
We used the data of the three radio sources jointly to obtain these
solutions.
The results are shown on Table~\ref{angular-separations-phx} and the
postfit residuals of the differenced phase delays 
in Fig.\ \ref{fig-diff-x}.

%
%
\begin{table}[htbp]
\caption{Relative right ascensions and declinations in J2000.0 coordinates of
the radio sources \object{BL\,1803+784}, \object{QSO\,1928+738} and \object{BL\,2007+777},
labeled as A, B, and C, respectively,
obtained by a weighted-least-squares analysis of the connected
phase delays. 
\label{angular-separations-phx}}
\begin{flushleft}
\begin{tabular}{lcr@{}r@{}r@{}l@{$\pm$}l} \hline
\multicolumn{7}{c}{{\bf Angular separations$^{\rm a}$}} \\ \hline
A--B & $\Delta \alpha$ &--1$^h$     & 27$^m$ & 2  & $\rlap{.}^s$811256 & 0$\rlap{.}^s$000062 \\
                     & $\Delta \delta$ &  4$^\circ$ & 30$'$  & 2  & $\rlap{.}''$44833  & 0$\rlap{.}''$00012 \\ \hline
B--C & $\Delta \alpha$ &--0$^h$     & 37$^m$ & 42 & $\rlap{.}^s$503305 & 0$\rlap{.}^s$000033 \\
                     & $\Delta \delta$ &--3$^\circ$ & 54$'$  & 41 & $\rlap{.}''$67756  & 0$\rlap{.}''$00013 \\ \hline
C--A & $\Delta \alpha$ &  2$^h$     & 4$^m$  & 45 & $\rlap{.}^s$314561 & 0$\rlap{.}^s$000052 \\
                     & $\Delta \delta$ &--0$^\circ$ & 35$'$  & 20 & $\rlap{.}''$77077  & 0$\rlap{.}''$00013 \\ \hline
\end{tabular}
\begin{list}{}{
\setlength{\leftmargin}{0pt}
\setlength{\rightmargin}{0pt}
}
\item[$^{\rm a}$]{Estimated errors are based on 
the combination of the various standard errors included in 
the analysis: statistical standard errors, and estimated standard errors in
the values of the (fixed) parameters of the theoretical model, in 
the ionospheric model from TEC measurements, and in the reference point 
chosen for each map.  See text and Tables~\ref{table-sens} and 
\ref{all-errors}.  Correlations among the estimated coordinates
for sources A and C, and among the sensitivities of those estimates to the
parameters listed in Table \ref{table-sens}, have 
been taken into account in computing the 
standard errors of the C-A separation.}
\end{list}
\end{flushleft}
\end{table}

We did not estimate in the weighted-least-squares analyses
the tropospheric zenith delays, the site coordinates, the 
coordinates of the reference source (i.e., \object{QSO\,1928+738};
using either of the other two sources as reference 
does not alter the results significantly), the coordinates of
the Earth's pole, and UT1-UTC. 
Instead, to accommodate software limitations we fixed 
their values
equal to their a priori values (see Table~\ref{table-model}), and
accounted for their a priori standard errors
in our subsequent sensitivity analysis
(see Table~\ref{table-sens}).
For each solution of the angular separations we estimated the
overall standard errors of the components of the
relative position between the radio sources as the 
root-sum-square (rss) of the standard deviations of
all the contributions discussed above 
(see Tables~\ref{table-sens} and \ref{all-errors}). 

%
%
\begin{table}[htbp]
\caption{Contributions $\delta\Delta\alpha$ and $\delta\Delta\delta$ to
the standard errors of the estimates of the sky coordinates of 
\object{QSO\,1928+738} relative to those of each of the two other radio sources 
obtained from the error sensitivity study.\label{table-sens}}
\begin{flushleft}
\begin{tabular}{l@{\,\,}cr@{\,}l@{\,\,}r@{\,\,}r@{\,\,}r@{\,\,}r} \hline
\multicolumn{8}{c}{{\bf Sensitivity analysis results$^{\rm a}$}}\\ \hline
                 &   &      &      & \multicolumn{2}{c}{{1803+784}} 
					   & \multicolumn{2}{c}{{2007+777}} \\
                 &   & \multicolumn{2}{l}{\footnotesize Standard}
				   & $\delta\Delta\alpha$
                                                              & $\delta\Delta\delta$
                                                                           & $\delta\Delta\alpha$
                                                                               &        $\delta\Delta\delta$ \\ 
{\footnotesize Parameter}       
		 &   & \multicolumn{2}{l}{\footnotesize deviation}  
                                   & [$\mu$as]$^{\rm b}$
						       & [$\mu$as] & [$\mu$as]$^{\rm b}$
										    & [$\mu$as] \\ \hline
{\footnotesize
Coordinates}  & $\alpha$
		     &  145 &$\mu$as$^{\rm b}$&     30 &      53 &      35 &      24 \\
{\footnotesize
of {1928+738}$^{\rm c}$} 
                 & $\delta$
	     	     & 280 &$\mu$as&     96 &      20 &      47 &      12 \\ \hline
{\footnotesize
Tropospheric}    & B & 0.1   & ns  &    105 &      20 &      22 &     105 \\
{\footnotesize
delay}           & F & 0.1   & ns  &     48 &      27 &      22 &      14 \\
{\footnotesize
at the zenith}   & P & 0.1   & ns  &     30 &      22 &      22 &       6 \\
{\footnotesize
of the radio}   & T & 0.1   & ns  &     12 &       9 &      22 &       7 \\
{\footnotesize
telescopes}      & X & 0.1   & ns  &     27 &      23 &      19 &       7 \\ \hline
                 & B & 2     & cm  &     60 &      62 &      41 &      40 \\
{\footnotesize
Radio }         & F & 2     & cm  &     27 &      18 &      16 &      16 \\
{\footnotesize
telescope   }    & P & 2     & cm  &     9  &      16 &      13 &      12 \\
{\footnotesize
coordinates$^{\rm c}$}    
                 & T & 2     & cm  &     12 &      18 &       9 &      10 \\
                 & X & 2     & cm  &     33 &      15 &      13 &      13 \\ \hline
{\footnotesize
Earth 	}        & x & 1     & mas &      3 &       2 &       0 &       1 \\
{\footnotesize
pole$^{\rm c}$}            
                 & y & 1     & mas &      0 &       0 &       0 &       1 \\ \hline
{\footnotesize
UT1-UTC$^{\rm c}$}         
                 &   & 0.04  & ms  &     33 &      60 &      38 &      27 \\ \hline \hline
{\footnotesize
rss$^{\rm d}$~
(estimated}      &   &       &     &        &         &         &          \\ 
{\footnotesize
standard error)} &   &       &     &{\bf 179}&{\bf 118}&{\bf 98}&{\bf 123} \\ \hline
\end{tabular}
\begin{list}{}{
\setlength{\leftmargin}{0pt}
\setlength{\rightmargin}{0pt}
}
\item[$^{\rm a}$]{Each entry in the last four columns was obtained by altering 
the a priori value of the parameter
in turn by one standard deviation and
repeating the least-squares analysis with the resultant table entry
being the absolute value of the difference between each result 
and the ``nominal" one.}
\item[$^{\rm b}$]{We use the factor 15$\cdot \cos \delta$ 
to convert $\mu$s to $\mu$as, 
where $\delta$ is the declination of the radio source.}
\item[$^{\rm c}$]{Standard deviations are twice the values published in
IERS (\cite{ier94}) (see Table~\ref{table-model}).}
\item[$^{\rm d}$]{The root-sum-square (rss) of the values in the 
corresponding columns.}
\end{list}
\end{flushleft}
\end{table}

%
%
%
\begin{table*}[htbp]
\caption{Error contributions to the three radio source positions
obtained from phase delays.
\label{all-errors}}
\begin{flushleft}
\begin{tabular}{p{7.5cm}rrrrrr} \hline
                  & \multicolumn{2}{c}{{1803+784}} 
                                  & \multicolumn{2}{c}{{1928+738}} 
					                      & \multicolumn{2}{c}{{2007+777}} \\
                  & $\delta\Delta\alpha$     
                                                              & $\delta\Delta\delta$
                          & $\delta\Delta\alpha$     
                                                              & $\delta\Delta\delta$
                                                                           & $\delta\Delta\alpha$
                                                                               & $\delta\Delta\delta$ \\ 
{\footnotesize Cause}
	      &  {\footnotesize [$\mu$as]$^{\rm a}$}
		            & {\footnotesize [$\mu$as]} 
                                  &  {\footnotesize [$\mu$as]$^{\rm a}$}
							    & {\footnotesize [$\mu$as]}      
					                                       & {\footnotesize [$\mu$as]$^{\rm a}$} 
				      	                                             & {\footnotesize [$\mu$as]} \\ \hline
{\footnotesize
Statistical standard error}
                 &  12 &  11 
					       & --$^{\rm b}$
							    & --$^{\rm b}$ 
									 &  12 &   9 \\
{\footnotesize
Fixed parameters in astrometric 
model$^{\rm c}$} 
                 & 179 & 118 
					       & --$^{\rm b}$ 
							    & --$^{\rm b}$ 
									 &  98 & 123  \\
{\footnotesize
Reference point in the map$^{\rm d}$}
                 &  14 &  14 &   23 &         12 &  23 &  12 \\
{\footnotesize
TEC$^{\rm e}$}
		 &  42 &  21
                                               & --$^{\rm b}$ 
                                                            & --$^{\rm b}$ &  16 &  14  \\ \hline
{\footnotesize
rss (estimated standard error)}           
                 & {\bf 186}
			    & {\bf 121} 
					       & {\bf 23}
							    & {\bf 12}
									       & {\bf 103} 
										    & {\bf 125} \\ \hline
\end{tabular}
\begin{list}{}{
\setlength{\leftmargin}{0pt}
\setlength{\rightmargin}{0pt}
}
\item[$^{\rm a}$] We use the factor 15$\cdot \cos \delta$
to convert $\mu$s to $\mu$as,
where $\delta$ is the declination of the radio source.
\item[$^{\rm b}$] We took \object{QSO\,1928+738} as reference
and estimated the positions of the other two sources
relative to the position of this reference.
\item[$^{\rm c}$] See Table 3.
\item[$^{\rm d}$] Each tabulated value includes the estimated
standard error in determining the peak of brightness due to the noise
in the map and
an additional contribution equal in magnitude to the differences between the
location of this peak and that of the centroid of the brightest
pixel (see text).
\item[$^{\rm e}$] The tabulated values are deduced from a sensitivity 
analysis for an estimated TEC standard error of
1.5$\times$10$^{16}$\,el/m$^2$ at each site; we assume that the TEC 
errors for each site are uncorrelated with those for the other sites.
\end{list}
\end{flushleft}
\end{table*}

We define ``sky closure" as the sums of the 
components of the
separation vectors between the three radio sources determined
pairwise. For right ascension $\alpha$, we have: 
${\cal C}_\alpha \equiv
\sum \Delta \alpha =
\Delta \alpha_{(C~-~B)}+
\Delta \alpha_{(B~-~A)}+
\Delta \alpha_{(A~-~C)}$,
where A, B, and C stand, 
respectively, for \object{BL\,1803+784}, 
\object{QSO\,1928+738}, and \object{BL\,2007+777}.  Closure in declination
$\delta$ is defined correspondingly.
The results of the sky-closure test are ${\cal C}_\alpha=-66\pm360$\,$\mu$as 
and ${\cal C}_\delta=15\pm220$\,$\mu$as, that is, zero to well within
the indicated standard errors.  We estimated these errors as the 
root-sum-square of the 
standard errors of the determinations of the pairwise separations.  
Consequently, these
sky closure errors are upper bounds since we ignored the correlations 
among the pairwise estimates of separation. 

%
%

\section{The pair \object{QSO\,1928+738}/\object{BL\,2007+777}: registrations and proper
motions\label{pair-1928-2007-registrations}}

We have obtained the pairwise separations among the radio
sources \object{BL\,1803+784}, \object{QSO\,1928+738}, and \object{BL\,2007+777} from observations
taken on 20-21 November 1991 through a weighted-least-squares
analysis of undifferenced
and differenced, connected, nearly ionosphere- and structure-free 
phase-delays, $\tau_\phi$.  

\begin{figure}[htbp]
\vspace{80mm}
\includegraphics{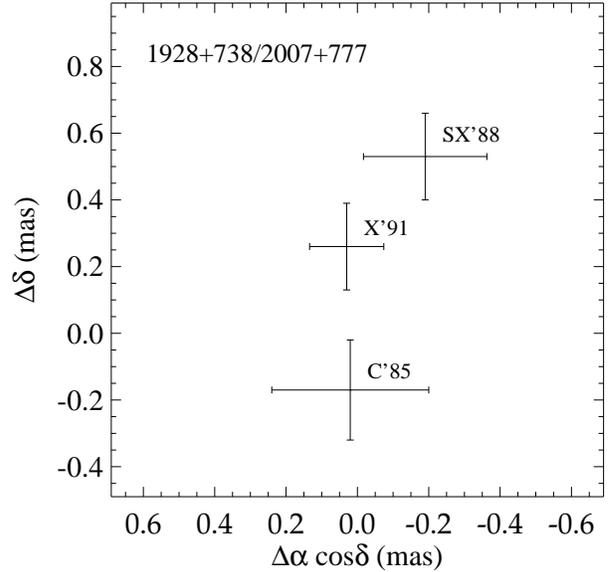}
\caption{The estimates of
the coordinates of \object{QSO\,1928+738} minus those of \object{BL\,2007+777}
(see Table 5).  The origin is chosen as in 
Table~5. 
Points are labeled to indicate the frequency band(s) used for
phase-delay astrometry and the year of each epoch.  S corresponds 
to 2.3\,GHz, C to
5\,GHz, and X to 8.4\,GHz.  SX indicates that both S and
X bands were used simultaneously.
\label{comp-19-20}}
\end{figure}

Considering previous observations in addition to ours,
we investigated apparent changes in the relative positions of 
brightness features in the sources \object{QSO\,1928+738} and \object{BL\,2007+777}. 
These radio sources were previously observed astrometrically 
in 1985.77 (Guirado \et\ \cite{gui95a}) and 1988.83 (Guirado \et\
\cite{gui98}).  
For them, we can compare the estimated separations for
the three epochs.  This comparison is shown 
in Table~\ref{positions-19-20} and Fig.\ \ref{comp-19-20}.  
For each epoch, the reference
position for each source is the peak of brightness of the corresponding
map.  Our result from epoch 1991.89 lies between the results from 
the epochs 1985.77 and 1988.83.

%
%
\begin{table*}[htbp]
\caption{Estimates at each of three epochs of the J2000.0 coordinates,
$\Delta\alpha$ and $\Delta\delta$, of \object{QSO\,1928+738} minus 
those of \object{BL\,2007+777}$^{\rm a}$.
\label{positions-19-20}}
\begin{flushleft}
\begin{tabular}{llcr@{$\pm$}lr@{$\pm$}lr@{$\pm$}l} \hline
        &           & Frequency& \multicolumn{4}{c}{$\Delta\alpha + 37^m42\rlap{.}^s503330$$^{\rm a}$} 
                                                           & \multicolumn{2}{c}{$\Delta\delta + 3^\circ54'41\rlap{.}''67780$$^{\rm a}$} \\
Epoch   & Antennas$^{\rm b}$
		    & [GHz]    & \multicolumn{2}{c}{[ms]} 
				              & \multicolumn{2}{c}{[mas]$^{\rm c}$}
				                           & \multicolumn{2}{c}{[mas]}\\ \hline
1985.77 & B,L,F,K,O & 5        &  0.005&0.053 &  0.02&0.22 & ~~~~~--0.17&0.15 \\
1988.83 & B,L,F,M,t & 8.4\&2.3 &--0.045&0.042 &--0.19&0.17 &  0.53&0.13 \\
1991.89 & B,F,P,T,X & 8.4      &  0.008&0.025 &  0.03&0.10 &  0.26&0.13 \\ \hline
\end{tabular}
\begin{list}{}{
\setlength{\leftmargin}{0pt}
\setlength{\rightmargin}{0pt}
}
\item[$^{\rm a}$] 
Since we assign the shifts in the separation between the two
radio sources to \object{QSO\,1928+738}, 
we present here the results obtained estimating the position
of \object{QSO\,1928+738}, while that of \object{BL\,2007+777} is held fixed at
$\alpha = 20^{h} 05^{m}  30\rlap{.}^s998546$, 
$\delta = 77^{\circ} 52' 43\rlap{.}''24777$ (IERS \cite{ier94}).
The reference coordinates
are the same as in Guirado \et\ (\cite{gui98}).
\item[$^{\rm b}$] The symbols correspond
to the following antennas (with diameter and
location given in parentheses): B, Effelsberg (100\,m, Germany); L, Medicina
(32\,m, Italy); K, Haystack (37\,m, Massachusetts); O, Owens Valley (40\,m, 
California); M, DSS63 (70\,m, Spain);
F, Fort Davis (25\,m, Texas); P, Pie Town (25\,m, New 
Mexico); t, Onsala (20\,m, Sweden); T, Kitt Peak (25\,m, Arizona); 
X, Los Alamos (25\,m, New Mexico).
\item[$^{\rm c}$] To facilitate 
comparisons between the values of $\Delta \alpha$ and $\Delta \delta$
in the table, we used the factor $15\cdot \cos\delta$, where
$\delta$ is the declination of \object{QSO\,1928+738}, to 
convert ms to mas.
\end{list}
\end{flushleft}
\end{table*}

We can obtain some insight into this result by investigating the alignment
of the maps.
The jets of \object{QSO\,1928+738} and \object{BL\,2007+777} 
are oriented nearly orthogonally on the sky:
The jet of \object{QSO\,1928+738} is directed to 
the south and that of \object{BL\,2007+777} to the
west.  This near orthogonality of the structures 
was one of the reasons we selected this pair.
Indeed, if the cores of the radio sources are assumed
stationary as in the standard model,
the motions in right ascension and declination of the 
jet components can be studied without confusion, using only the
differenced astrometric results.  
Alternatively, any apparent motion between the cores that cannot be plausibly
assigned to some other cause must bring the standard model into question
and/or the identification of the core.
For example, the possible emission of a new, unresolved component along a jet 
direction could change the position of the peak of brightness of 
the corresponding radio source.

\begin{figure}[htbp]
\vspace{110mm}
\includegraphics{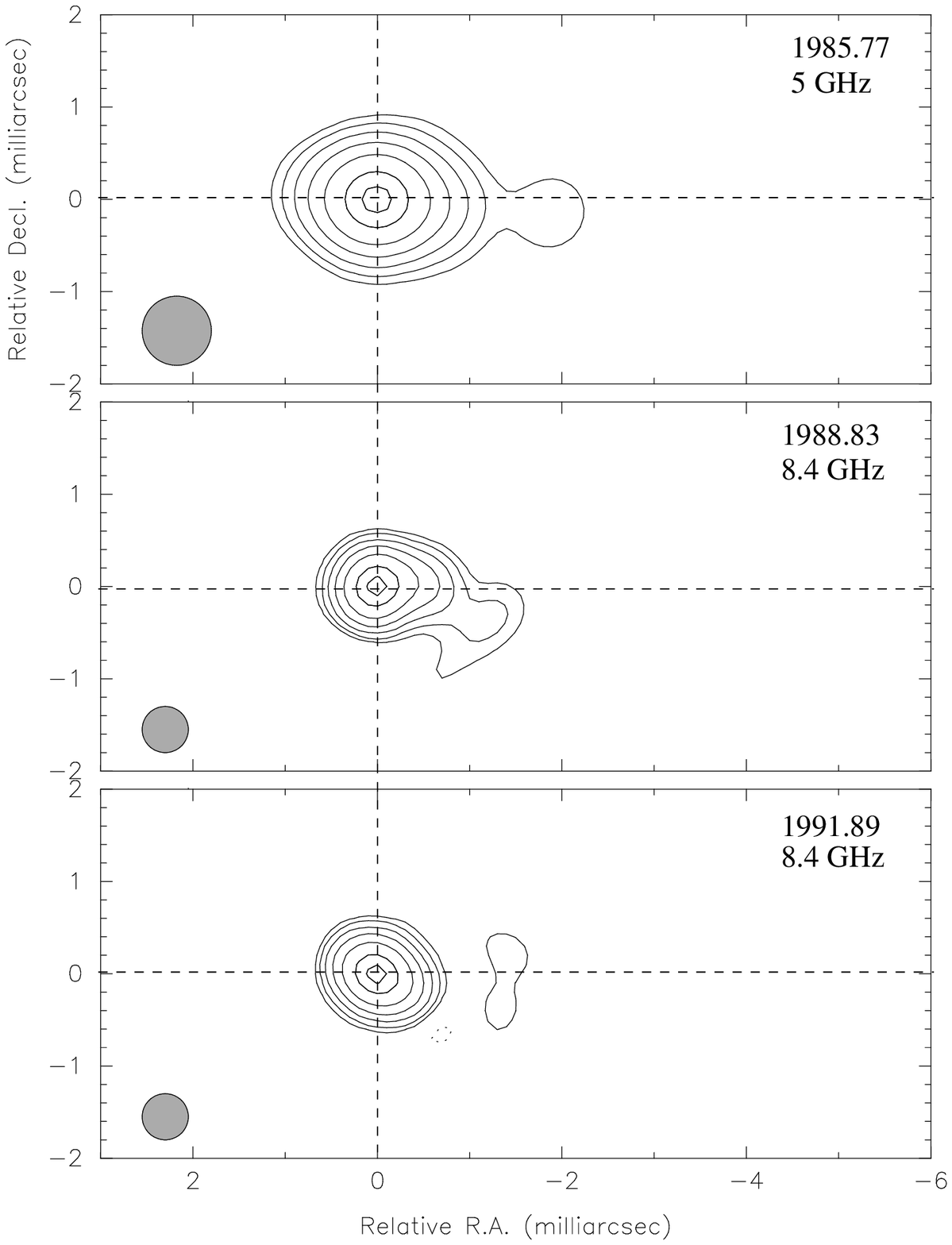}
\caption{Alignment of the maps of \object{BL\,2007+777}
based on the astrometric results.  Maps are overresolved and
contours represent from 2 to 90\% of the peak of brightness.  Convolving
beams (shown at left on each map) are circular, with diameters of
0.75, 0.5, and 0.5\,mas, top to bottom.
The crossings of the  dashed lines
show the reference points used in the astrometry.
The apparent differences in separation at different epochs 
between the two radio sources 
\object{QSO\,1928+783} and \object{BL\,2007+777} have been entirely 
attributed to \object{QSO\,1928+738} in this scenario (see text).
\label{20comp}}
\end{figure}

\begin{figure*}[htbp]
\vspace{186mm}
\includegraphics{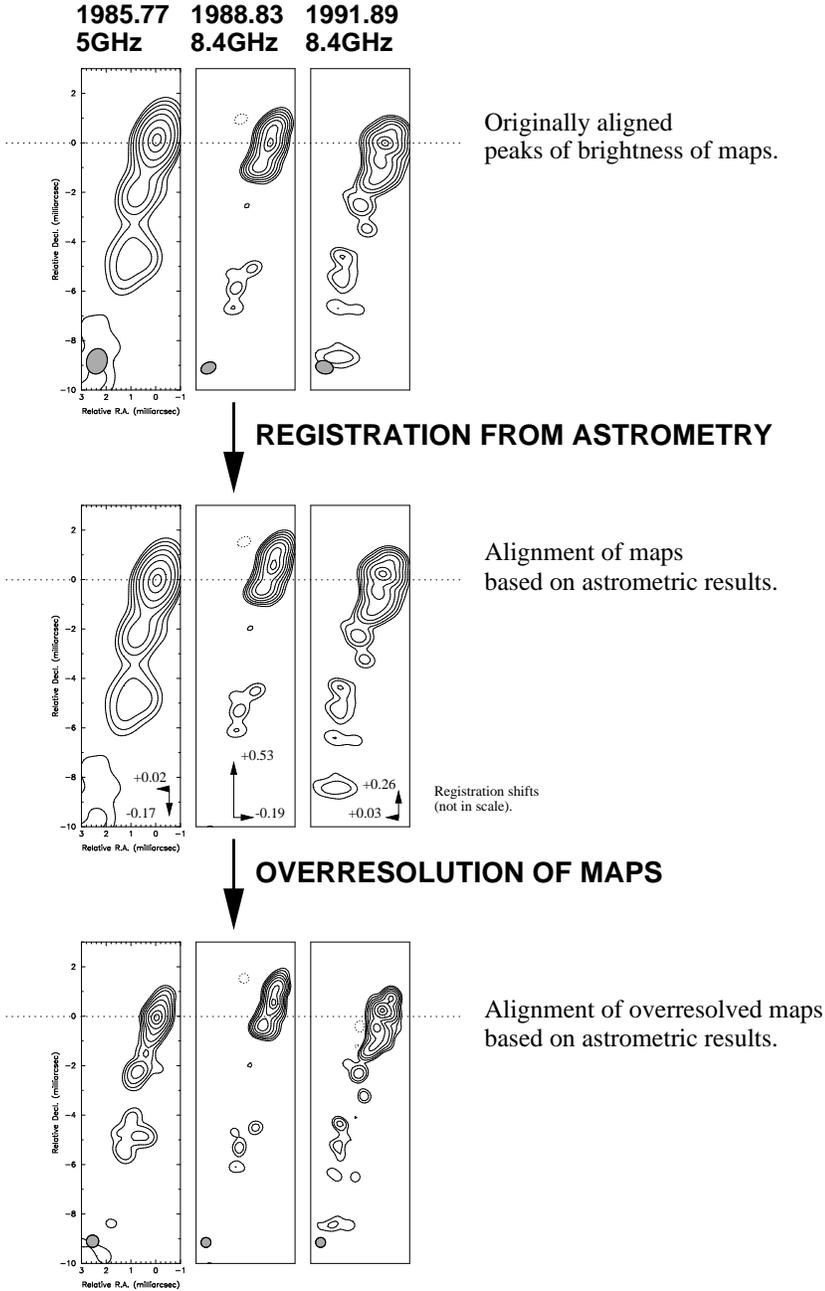}
\vspace{-6cm}
\hfill      \parbox[b]{5cm}{\caption[]{
Illustration of steps followed to try to properly
align the maps.
The astrometric shifts in milliarcseconds are represented (not to
scale) at the bottom of the middle figures.
The horizontal dotted line in the middle figures passes through the new
map origin.  The astrometric shifts are the displacements from the
origins originally chosen for the maps (upper figures) to the new
origins derived from astrometry (see Table 5).
\label{proces-regist-19}}
}
\end{figure*}

An ejection of a component in \object{QSO\,1928+738} might imply a
declination-oriented shift of the peak of brightness and
might affect the difference in declination between this radio source and
\object{BL\,2007+777}, but it would be unlikely to affect significantly the
difference in right ascension.
A similar analysis holds for \object{BL\,2007+777}.
Therefore, we interpret the predominantly north-south changes (see
Table \ref{positions-19-20}) in the separation 
of \object{QSO\,1928+738} and \object{BL\,2007+777} 
between these epochs as shifts with respect to the
chosen reference position which are
caused by changes in the structure 
of \object{QSO\,1928+738}.  
In other words, the chosen reference position is not the
(center of the) true core. 
In this scenario, the maps of \object{QSO\,1928+738}
are responsible for all changes: shifts of --170, 530, and 
260\,$\mu$as in declination, and 20, $-190$, and
30\,$\mu$as in right ascension for epochs 1985.77, 1988.83, and
1991.89, respectively. 
Accordingly, the corresponding alignment of 
the maps of \object{BL\,2007+777} (Fig.\ \ref{20comp}) shows no change in 
the position of the chosen
reference point.  The maps have been overresolved
to show more sharply the reference points chosen, which have been
placed at the origins of the maps.  We show in 
Fig.\ \ref{proces-regist-19} the steps followed to align the maps 
of \object{QSO\,1928+738} and the resulting registration.  
On the top, we have the three maps
aligned at the peaks of brightness as originally determined.
In the middle we show the result of shifting those
maps based on the results from astrometry and the assumption of no core motion
in \object{BL\,2007+777}.
At the bottom, we show the same alignment but with the images
overresolved to sharpen the distinction between the components, albeit
at the risk of being misled by the overresolution.

The remarkable lack of flux density from the northernmost
part of the maps from the first and third epochs, compared to that from the 
second is hard to understand and represents a clear puzzle in this scenario.
Further, it is difficult to understand the proposed
shift, with respect to the relative-position reference 
shown in Table~\ref{positions-19-20},
in the alignment (Fig.\ \ref{proces-regist-19})
of the maps of \object{QSO\,1928+738} from 1985.77 and 1988.83
in terms of a 
motion of particular features in the map (Guirado \et\ \cite{gui98}), 
another weakness in this scenario. 
Nevertheless, following through, we compare these 
two maps with that from 1991.89 and conclude tentatively that at each
of the three epochs the core is centered
about 1.5--2\,mas north of the previously chosen reference point.
This new identification of the core corresponds to the northernmost feature
of the 1988.83 map (undetected in this scenario in 1985.77 and 1991.89).

The core of \object{QSO\,1928+738} seems to be continually ejecting components, at
an estimated mean rate of one every 1.6 years (\cite{hum92}).   
The shifts displayed in
Fig.\ \ref{comp-19-20} are correspondingly
``explained" in terms of motions of the peak of brightness,
due to newly emerging components.  
In this scenario, we assume that our originally selected 
reference point corresponded to a different jet component in each map.
The changes in the flux
distribution over six years, a period in which, according to \cite{hum92}, 
up to four components are likely to have emerged, might well have caused
the peak of brightness to have moved during this period.
For the 1985.77 map at 5\,GHz, a difference in the location of the peak of
brightness and that of the
core of the radio source could also be due in part
to opacity effects, but with a value for the shift likely to be less than 
0.2\,mas
(applying a dependence with $k\lambda^\beta$
(Marcaide \et\ \cite{mar85}), with $0.7 < \beta < 2$, and $k$ corresponding
to 0.7\,mas for $\lambda$3.6--13\,cm, we obtain 0.10\,mas ($\beta=2$) to
0.21\,mas ($\beta=0.7$) for $\lambda$3.6--6\,cm).
This upper limit is small compared with the apparent astrometric shift.

A comparison of this Fig.\ \ref{comp-19-20} alignment with the
results provided by \cite{hum92} provides new insights into the problem.
\cite{hum92} identified the core of \object{QSO\,1928+738} as the 
northernmost (and most compact) component of the maps 
at 22\,GHz, based on a 5 year monitoring program.  These authors 
reported that this core had a flat spectrum, because of the similar
8.4 and 22\,GHz results.
Thus \cite{hum92} aligned their maps with this core
to determine the proper motions of the jet components that they reported.
They also developed a 
model for the kinematics of the jet components 
involving superluminal motion along a southward traveling sinusoidally 
curved jet ridge line.  In the model this path results from a 2.9\,year
period of precession of a binary black hole system that would produce a
sine function in the maps with a ``wavelength" of about 1.06\,mas and 
an ``amplitude" of about 0.09\,mas.
This predicted amplitude (in right ascension) is of
the same order of magnitude as the standard error of
our astrometric results.  The set of components
in the \cite{hum92} model
defines a ``traveling sine wave" moving at
0.28$\pm$0.05\,mas/yr, with no significant motions of the components with
respect to that of the sine wave.

\begin{figure*}[htbp]
\vspace{105mm}
\includegraphics{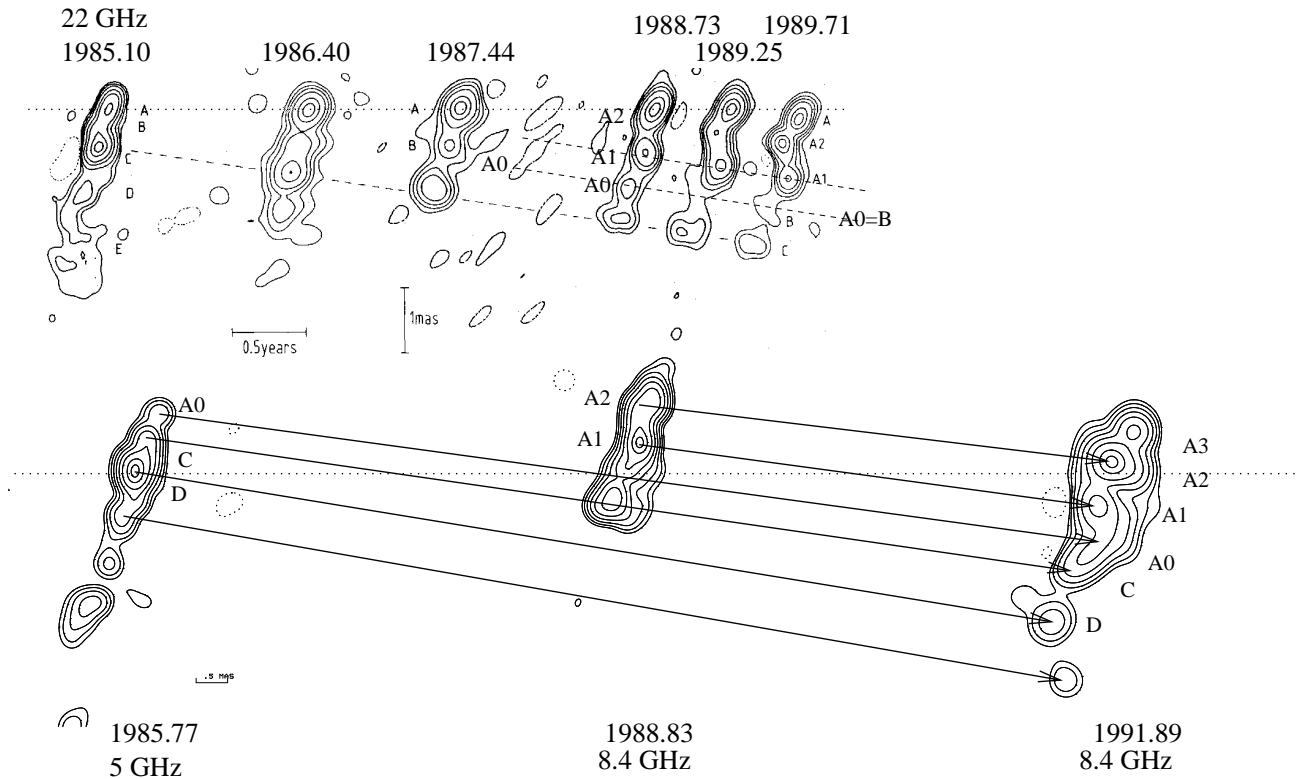}
\caption{Comparison, on the same angular size and temporal scales, 
of our results and those of H92 on \object{QSO\,1928+738}.
At the top, maps from H92's 22\,GHz observations
restored with circular restoring beams of 0.25, 0.30, 0.30, 0.25,
0.25, and 0.25 mas diameter, respectively
(left to right).  At the bottom, maps
from our alignment, with restoring
beams of 0.3, 0.3, and 0.4\,mas 
diameter (left to right). We propose identifications of 
corresponding features and components: For example, H92's C 
can be identified at 1985.77 and 1991.89.
The maps at 1988.73 (this paper) and 1988.83 (H92) 
show similar features.
We also propose identifications of components A1 and A2
in both sets of maps.  The dashed and solid lines indicate the proper
motions of the components.  The horizontal (dotted) lines are drawn 
for reference.
\label{comp-hummel}}
\end{figure*}

Our observations do not reach the resolution obtained by \cite{hum92},
due primarily to our almost threefold lower radio frequency.
We therefore cannot discern the inner structure of the radio sources 
in as much detail as \cite{hum92},
but we have been able to follow the changes in location of the
brightest features by means of our astrometry.
By overresolving our map and the maps of
Guirado \et\ (\cite{gui98}), we tried to identify
the components previously reported by
\cite{hum92}.
Specifically, we inferred that four jet
components were emitted between our first and third 
sets of observations, consistent with the results and predictions of 
\cite{hum92}.
These four components are labeled from south to north in 
Fig.\,\ref{comp-hummel} as A0, A1, A2, and A3.

%
%

A0 is identified tentatively as the oldest of these new
components (this component was labeled B by \cite{hum92},
as indicated in the upper part of Fig.\ \ref{comp-hummel}).  A0 can also
be identified in the maps of 1988.83 and 1991.89
in Fig.\ \ref{comp-hummel}.  It is fainter in these two maps than in
the  one of 1985.77.  Our estimate of its proper motion, based on
our conjectured astrometric alignment, is 
0.34$\pm$0.05\,mas/yr, with the quoted error being
the statistical standard error  (see Table\,\ref{tab-propmot}).

\begin{table}[htbp]
\caption{Position$^{\rm a}$ and proper motions of components in 
the maps of \object{QSO\,1928+738}.\label{tab-propmot}}
\begin{flushleft}
\begin{tabular}{crrrc}\hline
   & \multicolumn{3}{c}{Epoch}                         & \\
   & 1985.77         & 1988.83         & 1991.89         & Proper \\
   & 5\,GHz          & 8.4\,GHz        & 8.4\,GHz        & motion \\
   & [mas]           & [mas]           & [mas]           & [mas/yr] \\ \hline
A2 &      ---~~~~~   & --1.09$\pm$0.10 & --0.25$\pm$0.10 & 0.27$\pm$0.05 \\
A1 &      ---~~~~~   & --0.53$\pm$0.10 &   0.45$\pm$0.15 & 0.32$\pm$0.05 \\
A0 & --1.12$\pm$0.30 & --0.06$\pm$0.15 &   0.98$\pm$0.15 & 0.34$\pm$0.05 \\
C  & --0.49$\pm$0.15 &   0.44$\pm$0.10 &   1.43$\pm$0.15 & 0.32$\pm$0.03 \\
D  &   0.08$\pm$0.15 &      ---~~~~~   &   2.31$\pm$0.15 & 0.36$\pm$0.04 \\ \hline
\end{tabular}
\begin{list}{}{
\setlength{\leftmargin}{0pt}
\setlength{\rightmargin}{0pt}
}
\item[$^{\rm a}$] Angular distances of components are given with 
respect to the origin chosen, based on the astrometric alignment of the maps 
(see text and Fig.\ 7).
\end{list}
\end{flushleft}
\end{table}

A1 is a component that emerged between 1985.77 and 1988.83.  It
corresponds to the peak of brightness in the map of 1988.83 and
can also be identified in the 1991.89 map, yielding an estimated
proper motion of 0.32$\pm$0.05\,mas/yr (see
Table\,\ref{tab-propmot}).

A2 emerged later and represents the peak of bright\-ness in the
1991.89 map; in 1988.83 it had been near our initially assumed position
of the core.  Our estimate of its proper motion is
0.27$\pm$0.05\,mas/yr (see Table\,\ref{tab-propmot}).

The newest component, A3, apparently emerged after the 1988.83
epoch.  It is, we believe, south of the position of the (undetected) core, 
but, of course, north of the rest of the structure.

Component C marked in the \cite{hum92} maps can be identified in our maps
at 5 and 8.4\,GHz.  The different relative brightnesses of components C
and D in the 22 and 5\,GHz maps of 1985 might be due to C having still been
self-absorbed at 5\,GHz while
D was already in the optically thin phase of its evolution.
Our estimate of the proper motion of component C from the three maps 
is 0.32$\pm$0.03\,mas/yr
(see Table\,\ref{tab-propmot}).

It is remarkable that D is not visible in the 1988.83 map, even though
it is the brightest feature of the 1985.77 image. Similarly,
C is at least twofold brighter and also more extended in the 
1988.83 map than we would have expected from its evolution along a jet.  
The brightness change and the
extension change can 
perhaps be explained by the undersampling
of the $(u,v)$ plane for intermediate spatial frequencies.
Table\,\ref{positions-19-20} shows that the array at this epoch consisted
of four European radio telescopes and one in the U.S. (Fort
Davis).  This array provided sampling only at 30-50\,M$\lambda$
for six baselines and at about 200\,M$\lambda$ for the four intercontinental
ones.  Intermediate spatial frequencies were not sampled.
Hybrid mapping could have assigned to the inner features
the flux density of these undetected components, as model fitting confirms.  
Specifically, for this
poor sampling of those intermediate spatial frequencies,
the visibilities of the map from Guirado \et\ (\cite{gui98}) 
are essentially equivalent 
to those generated by a model, consisting of elliptical Gaussian
components fit to the observed visibility data, that includes a component
located where 
we believe
component D is located.  
The other components
of the model were located as found in the hybrid maps.
The proper motion of component D, deduced from only our first
and third epochs, is 0.36$\pm$0.04\,mas/yr (statistical standard error).

Our proper motion results agree with those reported by
\cite{hum92}. Although we have lower resolution data than they,
we took advantage of our ability to align the maps of \object{QSO\,1928+738} 
based on our accurate astrometric results, to obtain a longer time base.

%
%

\section{Conclusions\label{conclusions}}

We have shown that at 8.4\,GHz phase delays can be ``connected"
reliably for sources 7$^\circ$ apart on the sky even at epochs 
of maximum solar activity.
Guirado \et\ (\cite{gui95a}) and Lara \et\ (\cite{lar96}) demonstrated
such phase connection for angular distances of
$\sim$5$^\circ$, but not for epochs of high solar activity and
rapidly varying TEC.
Here, under such conditions, we succeeded at phase 
connection with 8.4\,GHz, but not with 2.3\,GHz data.

We removed most of the structure phase from the astrometric
observables based on the hybrid maps obtained from
our observations and the selection of reference points in the maps, as
in the past.  Nevertheless, errors at the milliarcsecond 
level in the comparison of astrometric positions
determined from data obtained at different epochs can result from
changes in the structures of the radio sources, if such changes
are not taken into account properly.
We try to diminish such error or bias, by attempting to remove
the effects of the source structure from
our data through referring all astrometric measurements to the 
core (i.e., the same ``physical" point) in the source.

As a possible improvement to the general astrometric
technique, we removed most of the effect of
the ionosphere by means of TEC estimates deduced from GPS
measurements at stations in the vicinity of the radio telescopes that we used.
This important step forward is vital when adequate dual-band VLBI observations
are not available.
Moreover, given sufficiently accurate corrections for ionospheric effects 
from GPS measurements, only single-band VLBI observations need to be performed
so that higher sensitivity (e.g., for fainter sources) can be obtained
by using the whole recording bandwidth for this single band.
The technical aspects of the procedure will be presented in a complementary
paper (Ros \et\ \cite{ros99}).

After obtaining the pairwise relative positions of the three radio sources, we
compared the separation of the pair \object{QSO\,1928+738}/\object{BL\,2007+777} 
with two previous results from earlier epochs.
Our relative position fell between the other two: 
we interpreted the differences as being due to different ``physical"
reference points in the sources being used (unintentionally) 
at the different epochs of observation.

According to \cite{hum92},
\object{QSO\,1928+738} is a superluminal radio source that
ejects components which seem to move along a 
sinusoidal trajectory
starting at a flat-spectrum component detected 
at 22\,GHz that \cite{hum92} identified with the core.
However, the inferences drawn from what seemed to be the correct 
alignment of our maps, led us to question
this identification, suggesting instead that the core is further north.
We then identified four components that emerged during the six years 
spanned by our observations (late 1985 to late 1991), 
which we
labeled A0, A1, A2, and A3; their proper motions all appear to be in the
range of 0.3-0.4\,mas/yr.

In the future, the improvements incorporated into our present data
reduction can perhaps be successfully applied to radio sources much 
further apart on the sky and at any part of the solar activity cycle
with the use of only single-band observations.
In addition, the better performance in recent years of the global VLBI 
network (especially the VLBA) provides sizable advantages for
phase-delay astrometry: higher slew speeds of radio
telescopes with corresponding reduction of slew times, 
better sensitivity of receivers, and automated data correlation with the
corresponding reduction of ``waiting" times.  We believe that far
more ambitious projects are now possible.

\begin{acknowledgements}
We wish to thank the staffs of all the observatories for their
contributions to the observations, especially J.\ Ball and C.\ Lonsdale
of the Haystack Observatory.  We also thank
in particular the MPIfR staff for their efforts
during the correlation.
We want to acknowledge P.\ El\'osegui for his valuable help during
the data collection; A.\ Alberdi,
S.\ Britzen, 
A.M.\ Gontier,
L.\ Lara,
and 
K.J.\ Standke 
for their help during calibration and hybrid mapping;
and E.\ Sard\'on for her valuable help with 
the ionospheric analysis.  GPS data were provided by the GARNER archive
at SOPAC, University of California-San Diego.
E.R.\ acknowledges an F.P.I.\ Fellowship of the Generalitat
Valenciana and a Comett grant of the E.U.\ to ADEIT for a stay at
MPIfR.  This work has been partially supported by the Spanish DGICYT
grants PB\,89-0009, PB\,93-0030, and PB\,96-0782, and by the 
U.S.\ National Science Foundation Grant No.\ AST 89-02087.
\end{acknowledgements}



\end{document}